\documentclass[default]{aastex701}
\shorttitle{Compact IR source in Cyg X-3}
\shortauthors{Igor I.Antokhin}
\received{...}
\revised{...}
\accepted{...}
\submitjournal{RNAAS}
\begin{document}

\title{On the location of the compact infrared source in Cyg X-3}

\author[orcid=0000-0002-3561-8148,sname='Antokhin']{Igor I. Antokhin}
\affiliation{Lomonosov Moscow State University, \\
Sternberg State Astronomical Institute, \\
119992, Universitetskij prospect, 13, Moscow, Russian Federation}
\email[show]{igor@sai.msu.ru}

\begin{abstract}

This note complements the article on X-ray and infrared variability of the X-ray binary Cyg X-3 (WR+c), published by me and my co-authors \citep{ant22}. In that paper, a compact IR source was discovered in the system, located in the vicinity of the X-ray source associated with the relativistic companion. In the current note I refine the possible location of the IR source based on simple qualitative considerations.

\end{abstract}


%\section{Results of the previous work}\label{results}

\section{What do the X-ray and IR light curves of Cyg X-3 tell us about the system?}\label{results}

\begin{figure*}[t!]
\plotone{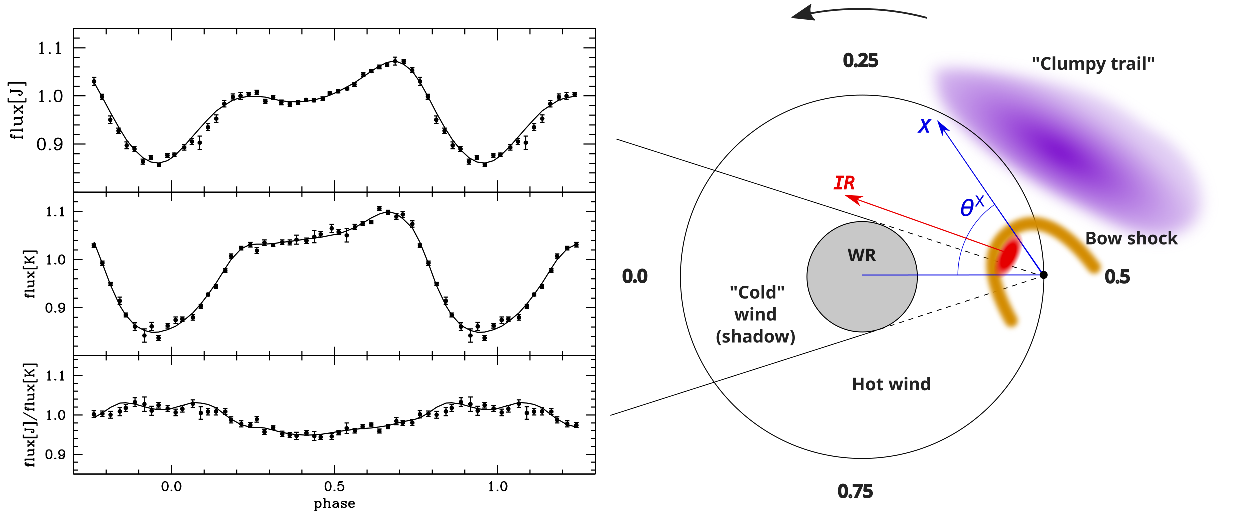}
\caption{Left: observational (dots) and theoretical (solid lines) {\em JK} light curves of Cyg X-3 in normalized fluxes, and the corresponding color curves. This is a slightly simplified Fig.\,11 from Paper I. Right: schematic representation of the system in our model from the same paper (Fig.\,7). The orbit is shown by the black circle. The relativistic companion is shown by the black dot. Blue solid line and arrow have been added to the original figure, showing the direction to the apex of the bow shock in the X-ray model and the corresponding angle $\theta^{X}$. The possible refined position of the compact IR source and the direction from it to the maximum IR absorption are shown by the red area and the red arrow.}
\label{fig:lc_model}
\end{figure*}

Cyg X-3 is the only known binary system in our Galaxy consisting of a WR star and a relativistic object \mbox{(WN4-8+c)}. In \cite{ant22} (Paper I) we analyzed the X-ray {\em RXTE} ASM and infrared (IR) {\em JHK} light curves of the system. Observational and model {\em JK} light/color curves are shown in Fig.\,\ref{fig:lc_model} on the left. This figure is a slightly simplified version of Fig.\,11 in Paper I. The observational IR light curves are qualitatively similar to the X-ray ones: in both cases the primary minimum shows fast ingress and slow egress, both X-ray and IR light curves show a decrease of the flux near the orbital phase $\sim 0.4$. In the IR, this decrease is much more pronounced than in the X-rays.

The X-ray source in the system is a compact formation: the accretion disk around the relativistic companion or the so-called ``focused wind''. The overall quasi-sinusoidal X-ray flux variations are obviously caused by the variable absorption of the X-ray source radiation in the wind of the WR star during its orbital revolution. A possible explanation for the X-ray absorption at the orbital phase $\sim 0.4$ was given by \cite{vilhu13}. The authors suggested that it could be caused by the absorption of the X-ray radiation by the so-called ``clumpy trail'', formed due to the interaction of the WR star wind with the relativistic jet. The asymmetry of the ingress and egress in the case of WR+c binaries can be explained by the absorption of the X-ray radiation in the bow shock in front of the relativistic companion (see Paper I for details). The bow shock is positioned in such a way that it creates additional absorption just at the orbital phases $0.1-0.3$, resulting in a slow egress.

In the IR domain, before Paper I, the overall variability was explained by the variable free–free emission of the two-component WR wind (a hot part illuminated by the X-ray source and a relatively cold part shadowed by the WR body \citep{vankerk93, vankerk96}. However, given the similarity between the observed X-ray and IR light curves of Cyg X-3 found in our work, the only reasonable explanation for such a similarity is the existence of a compact IR source located near the X-ray source. Then its radiation passes through the same absorbing structures as the X-ray radiation, resulting in a similar Ilight curve. This conclusion is further supported by the variability of the IR color of the system: at the orbital phase $\sim 0.0$ (the X-ray source behind WR), the color becomes bluer. In the free-free emission of the two-component WR wind model, the color should become redder. On the other hand, the compact IR source suggested above naturally explains the blue shift. Indeed, the free-free absorption is proportional to the square of wavelength, resulting in a blue shift in the color of the IR source and the entire system at the phase $\sim 0.0$.

A schematic view of the binary model used in Paper I is shown on the right in Fig.\,\ref{fig:lc_model}. In X-ray, the only radiation source was a point source located at the position of the relativistic companion. Its radiation along the line of sight at different orbital phases is absorbed in the hot part of the WR wind, in the bow shock and in the ``clumpy trail''. In IR, there were two radiation sources: a point source located at the same position as the X-ray source, and the free-free radiation from the two-component WR wind. The radiation from the compact source is absorbed in the same structures as the radiation from the X-ray source.

In Paper I, using the described model, various parameters of the system were determined: the orbital inclination, the directions to the apexes of the bow shock and clumpy trail, their characteristic angular sizes, the mass loss rate of the WR star, and so on. Most of these parameters were determined independently from the observed X-ray and IR light curves to check their consistency. It turned out that the X-ray and IR modeling results agree quite well, with the exception of one parameter: the direction to the apex of the bow shock. In X-ray, the angle between the system axis and the direction to the bow shock apex was $\theta^{X}=55\degr\pm2\degr$, in a very good agreement with the theoretical value of $55\fdg3$ for the orbital and WR wind parameters of Cyg X-3. However, in IR this angle was $\theta^{IR}=16\degr\pm3\degr$ and $22\degr\pm2\degr$ in {\em J} and {\em K} filters, respectively. This contradiction remained unexplained in Paper I. It was only after the paper had already been published that I realized there was a simple and natural explanation. I will present it in the next section.

\section{New considerations and conclusion}

In the model described above, we assumed that the point IR source was located at the same position as the point X-ray source. We also assumed that the absorption in the bow shock is symmetric about its apex. In reality, this may not be the case, and the densities of the two branches of the bow shock facing inward and outward of the orbit may differ (the former perhaps larger). In Paper I we discussed a possible nature of the compact IR source. One possibility was the formation of the IR emission in the dense inner parts of the bow shock (see the paper for details). If we place a relatively compact IR source somewhere in the inner part of the bow shock, the direction of the maximum IR absorption (which is what determined the angle $\theta^{IR}$ in our model) may change, becoming smaller. This is illustrated in Fig.\,\ref{fig:lc_model} on the right, where the possible location of the compact IR source and the direction of maximum IR absorption are shown in red.

Thus, the difference between the directions to the bow shock apex in X-ray and IR can be explained by the fact that in Paper I we assumed that the positions of the corresponding point sources coincide with each other and with the position of the relativistic companion. In reality, the difference in these directions may mean that the IR source is located at some distance from the X-ray source, presumably in the direction of the bow shock. I hope this information may be useful in the development of future Cyg X-3 models.

\begin{acknowledgments}

I thank the Russian Science Foundation for supporting my work through the grant 23-12-00092.

\end{acknowledgments}

\bibliography{Antokhin}{}

\begin{thebibliography}{}
\expandafter\ifx\csname natexlab\endcsname\relax\def\natexlab#1{#1}\fi
\providecommand{\url}[1]{\href{#1}{#1}}
\providecommand{\dodoi}[1]{doi:~\href{http://doi.org/#1}{\nolinkurl{#1}}}
\providecommand{\doeprint}[1]{\href{http://ascl.net/#1}{\nolinkurl{http://ascl.net/#1}}}
\providecommand{\doarXiv}[1]{\href{https://arxiv.org/abs/#1}{\nolinkurl{https://arxiv.org/abs/#1}}}

% type= article
\bibitem[{I.~I. {Antokhin} {et~al.}(2022){Antokhin}, {Cherepashchuk},
  {Antokhina}, \& {Tatarnikov}}]{ant22}
{Antokhin}, I.~I., {Cherepashchuk}, A.~M., {Antokhina}, E.~A., \& {Tatarnikov},
  A.~M. 2022, \bibinfo{title}{{Near-IR and X-Ray Variability of Cyg X-3:
  Evidence for a Compact IR Source and Complex Wind Structures},} \apj, 926,
  123, \dodoi{10.3847/1538-4357/ac4047}

% type= article
\bibitem[{M.~H. {van Kerkwijk}(1993){van Kerkwijk}}]{vankerk93}
{van Kerkwijk}, M.~H. 1993, \bibinfo{title}{{Spectroscopic and photometric
  variability of Cygnus X-3.},} \aap, 276, L9

% type= article
\bibitem[{M.~H. {van Kerkwijk} {et~al.}(1996){van Kerkwijk}, {Geballe}, {King},
  {van der Klis}, \& {van Paradijs}}]{vankerk96}
{van Kerkwijk}, M.~H., {Geballe}, T.~R., {King}, D.~L., {van der Klis}, M., \&
  {van Paradijs}, J. 1996, \bibinfo{title}{{The Wolf-Rayet counterpart of
  Cygnus X-3.},} \aap, 314, 521, \dodoi{10.48550/arXiv.astro-ph/9604100}

% type= article
\bibitem[{O. {Vilhu} \& D.~C. {Hannikainen}(2013){Vilhu} \&
  {Hannikainen}}]{vilhu13}
{Vilhu}, O., \& {Hannikainen}, D.~C. 2013, \bibinfo{title}{{Modeling the X-ray
  light curves of Cygnus X-3. Possible role of the jet},} \aap, 550, A48,
  \dodoi{10.1051/0004-6361/201219843}

\end{thebibliography}
\bibliographystyle{aasjournalv7}



\end{document}